\pdfoutput=1
\documentclass{webofc}
\usepackage[varg]{txfonts}   %
\usepackage{hyperref}
\usepackage{url}
\usepackage{booktabs}
\usepackage{array}
\RequirePackage{orcidlink}   %
\hypersetup{colorlinks=true,citecolor=blue,urlcolor=blue,linkcolor=blue,
  pdftitle={SPECTRUM: A Strategic Framework and Technical Blueprint for European Exascale Research Data and Compute Infrastructure},
  pdfsubject={CHEP 2026 proceedings. Funded by SPECTRUM, European Union Grant Agreement Number 101131550},
  pdfkeywords={SPECTRUM, Horizon Europe, grant agreement 101131550, compute and data continuum, HL-LHC, SKA, SRIDA, Technical Blueprint}}
\graphicspath{{figures/}}
\begin{document}
\AddToHookNext{shipout/foreground}{\put(0.5\paperwidth,-1.4cm){\makebox[0pt]{\parbox[t]{\textwidth}{\centering\footnotesize\itshape Preprint of a paper submitted to the proceedings of the 28th Conference on Computing in High Energy and Nuclear Physics (CHEP 2026), to be published in EPJ Web of Conferences.}}}}

\title{SPECTRUM: A Strategic Framework and Technical Blueprint for European Exascale Research Data and Compute Infrastructure}

\author{\firstname{Sergio} \lastname{Andreozzi}\inst{1}\orcidlink{0000-0001-5567-4000}\fnsep\thanks{\email{sergio.andreozzi@egi.eu}} \and
        \firstname{Stefano} \lastname{Bagnasco}\inst{2}\orcidlink{0000-0001-6062-6505} \and
        \firstname{Florian} \lastname{Berberich}\inst{3}\orcidlink{0000-0001-6695-9570} \and
        \firstname{Tommaso} \lastname{Boccali}\inst{4}\orcidlink{0000-0002-9930-9299} \and
        \firstname{Concezio} \lastname{Bozzi}\inst{5}\orcidlink{0000-0001-6782-3982} \and
        \firstname{Luis} \lastname{Cifuentes}\inst{3}\orcidlink{0000-0001-9077-2597} \and
        \firstname{Luca} \lastname{dell'Agnello}\inst{6}\orcidlink{0000-0003-1086-6740} \and
        \firstname{Ama\"el} \lastname{Ellien}\inst{7}\orcidlink{0000-0002-1038-3370} \and
        \firstname{Chiara} \lastname{Ferrari}\inst{7}\orcidlink{0000-0002-2917-9759} \and
        \firstname{Luciano} \lastname{Gaido}\inst{2}\orcidlink{0000-0002-6994-4988} \and
        \firstname{Maria} \lastname{Girone}\inst{8} \and
        \firstname{Hanno} \lastname{Holties}\inst{9}\orcidlink{0000-0002-8436-0534} \and
        \firstname{Hans-Christian} \lastname{Hoppe}\inst{3} \and
        \firstname{Valentin} \lastname{Jay~Blanchard}\inst{10}\orcidlink{0009-0007-9509-7237} \and
        \firstname{Corentin} \lastname{Lefevre}\inst{10}\orcidlink{0000-0002-2933-2370} \and
        \firstname{Kristen} \lastname{Lutz}\inst{11} \and
        \firstname{Andrea} \lastname{Manzi}\inst{1}\orcidlink{0000-0001-7949-2199} \and
        \firstname{Shan} \lastname{Mignot}\inst{7}\orcidlink{0000-0002-8515-939X} \and
        \firstname{Raymond} \lastname{Oonk}\inst{11} \and
        \firstname{Xavier} \lastname{Salazar Forn}\inst{1}\orcidlink{0009-0007-2076-199X} \and
        \firstname{David} \lastname{Southwick}\inst{8}\orcidlink{0000-0003-4229-7191} \and
        \firstname{John} \lastname{Swinbank}\inst{9}\orcidlink{0000-0001-9445-1846} \and
        \firstname{Jeff} \lastname{Wagg}\inst{7} \and
        \firstname{Eric} \lastname{Wulff}\inst{8}\orcidlink{0000-0002-4681-8516}
}

\institute{EGI Foundation, Amsterdam, Netherlands
\and INFN Sezione di Torino, Torino, Italy
\and Forschungszentrum J\"ulich, J\"ulich Supercomputing Centre, J\"ulich, Germany
\and INFN Sezione di Pisa, Pisa, Italy
\and INFN Sezione di Ferrara, Ferrara, Italy
\and INFN CNAF, Bologna, Italy
\and Observatoire de la C\^ote d'Azur, Universit\'e C\^ote d'Azur, CNRS, Nice, France
\and CERN, Geneva, Switzerland
\and ASTRON, Netherlands Institute for Radio Astronomy, Dwingeloo, Netherlands
\and NEOVIA Innovation, Paris, France
\and SURF B.V., Utrecht, Netherlands
}

\abstract{The HiLumi LHC, the SKA Observatory and LOFAR~2.0 will take High Energy Physics (HEP) and Radio Astronomy (RA) towards exabyte-scale data within the next decade, beyond the capacity of today's infrastructure. The SPECTRUM project responded with two documents built on the same evidence. A Technical Blueprint describes a European compute and data continuum as a capability map, consolidates the gaps found in HEP, RA and related use cases, and recommends technical actions in eight areas. A Strategic Research, Innovation and Deployment Agenda (SRIDA) organises thirteen priorities in four pillars, maps them to investment areas and European funding instruments, and phases them over a roadmap set against the timelines of the three instruments. This paper, by editors and contributors of both documents, summarises them, their consultative method and the impact the SRIDA expects.}

\maketitle

\section{Introduction}
\label{intro}

High Energy Physics (HEP) and Radio Astronomy (RA) enter the exascale era in the same decade. The LHC, where the Higgs boson was discovered, is undergoing a major upgrade. The High-Luminosity LHC (HiLumi LHC) should be operational in mid-2030, to study the Higgs boson in greater detail and look for rare new phenomena~\cite{hilumi}. The two largest LHC experiments, ATLAS and CMS, project disk needs of about 3~EB and 1.5~EB and tape archives of 4 to 6~EB by the mid-2030s~\cite{d61,atlasroadmap,cmsphase2}; ALICE expects its computing needs to grow by about 15\% a year until 2041, and LHCb its simulation output about 7.5-fold from Run~4 to Run~5~\cite{d51}. LOFAR~\cite{vanhaarlem} surveys the low-frequency radio sky to study how galaxies, clusters and active galactic nuclei form and evolve. Its upgrade, LOFAR~2.0, will start full operations during 2027; it is expected to need about 100~PB of archive storage over five years and 200 million core-hours of processing a year~\cite{d51}. The SKA Observatory~\cite{scaife}, whose science cases include the formation of structure in the early universe and neutral hydrogen in galaxies~\cite{d51}, processes its data streams in Science Data Processors near the telescope sites in Australia and South Africa~\cite{d53}. During full operations it will distribute up to 700~PB of observatory data products a year to the SKA Regional Centre Network (SRCNet)~\cite{d51,d61}. At full operation, from 2030 onwards, the HiLumi LHC and the SKA Observatory will each generate hundreds of petabytes a year, an order of magnitude beyond current capacity~\cite{d71}.

The large-scale computing, data and network capabilities that both communities use come from several sources. Central facilities, such as the CERN Tier-0 and the SKA Science Data Processors, are funded by the host organisation. Most other capacity is pledged by countries and institutes: WLCG rests on 65 memoranda of understanding signed by funding agencies~\cite{wlcgstrategy}, the LOFAR Long Term Archive already runs as a distributed system at three sites under service-level agreements~\cite{d53}, and SRCNet, still under construction, is developing comparable arrangements~\cite{d61,d71}. The pledged capacity is itself part of wider pan-European digital infrastructures: European WLCG Tier-1 centres, for example, belong to the EGI Federation's high-throughput infrastructure, which serves many other communities~\cite{d61}. The EGI Federation also supports WLCG with federation services~\cite{wlcgstrategy}, EOSC offers channels for the secondary reuse of research data, and G\'EANT connects the sites through its high-speed network. More recently, EuroHPC has offered access to its supercomputers, where HEP and RA can explore extending their workloads; its allocation model differs from pledges, granting time by competitive calls, typically for one year~\cite{d71}.

SPECTRUM, a Horizon Europe coordination and support action (January 2024 to June 2026), produced two documents on the compute and data continuum that Europe needs for HEP and RA, together with research infrastructures, digital infrastructure providers and policy stakeholders. The Technical Blueprint (D6.1)~\cite{d61} covers what the continuum must deliver, for infrastructure architects. The Strategic Research, Innovation and Deployment Agenda (SRIDA, D7.1)~\cite{d71} identifies who should act, when and with which resources, for programme owners, funders and infrastructure leaders. Compared with earlier studies, SPECTRUM adds an integrated analysis of HEP and RA requirements and gaps, with one template for both, joint actions mapped to European funding instruments and phased in a roadmap, and a proposal to credit users in units of environmental impact in place of core-hours. The SRIDA's priorities rest mainly on HEP and RA needs. For other communities to judge which findings transfer, it lists its assumptions: data correlated or reconstructed centrally, non-personal, open after time-limited embargoes, processed for throughput with real-time filtering at source, and governed by formal multinational structures over decades~\cite{d71}. It also names the network underpinning the continuum as important but leaves its detailed treatment to follow-up work.

\section{Method}
\label{sec-method}

\begin{figure}[t]
\centering
\includegraphics[width=\textwidth]{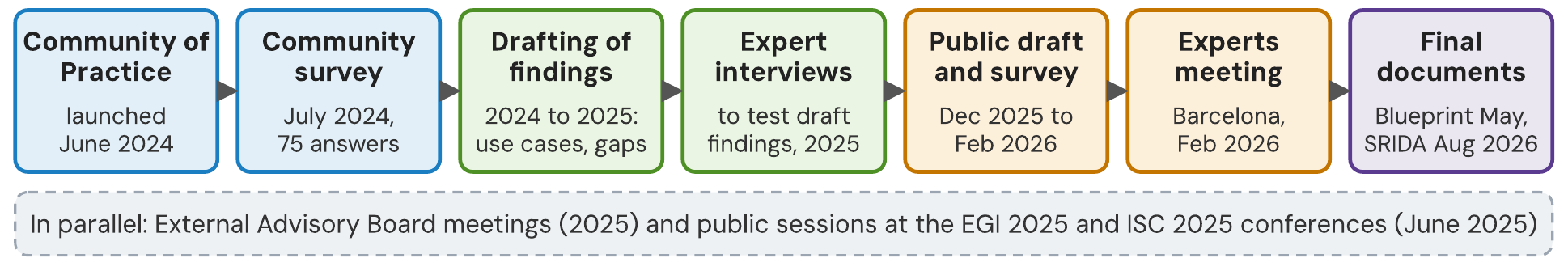}
\caption{The consultative process behind the Technical Blueprint and the SRIDA, 2024 to 2026, from D3.1, the Experts Meeting summary and D7.1 (Section~2, Annex~C and version history)~\cite{d31,expertsmeeting,d71}.}
\label{fig-process}
\end{figure}

Both documents came out of one consultative process (Figure~\ref{fig-process}). The Community of Practice, launched in June 2024, had 65 distinct participants from HEP, RA and digital infrastructures in six working groups at the start of 2025, and a survey of these communities in July 2024 received 75 valid answers, about a quarter from infrastructure managers~\cite{d31}. The findings were then drafted from fourteen use cases (six from HEP, five from RA and three from other domains) documented with one template~\cite{d51}, an assessment of twenty European digital infrastructures~\cite{d53}, an analysis of access policies~\cite{d52} and desk research on strategy documents and earlier studies~\cite{d71}. Among the earlier studies, the HEP Software Foundation roadmap set the software and computing R\&D agenda for the HiLumi LHC era~\cite{hsfcwp}; the WLCG Strategy 2024--2027 calls for a reference implementation for integrating high performance computing (HPC) and multi-year resource planning, and for WLCG to consider pledges from time-limited allocations~\cite{wlcgstrategy}; the Strategy for HPC Integration in WLCG/HEP (2025) sets out the technical, organisational and policy barriers to using HPC~\cite{girone2025}; and the JENA computing initiative, which ran a joint survey with SPECTRUM~\cite{d31}, made compatible recommendations on HPC access, interoperability and identity~\cite{jena2025}.

Working groups led by project partners drafted the SRIDA priorities, which were grouped into pillars in November 2025; interviews with community experts tested the draft findings, and the External Advisory Board advised on the priority framework (D7.1, Annex~C). Polls at the EGI 2025 and ISC 2025 conferences gave early public input. The drafts were published on 11 December 2025~\cite{drafts}; the consultation survey received 26 completed responses from nine countries (25 on the Blueprint, 10 on the SRIDA), and invited experts discussed the results in Barcelona on 18 and 19 February 2026~\cite{expertsmeeting}. The SRIDA's investment and roadmap sections were written after this consultation.

The SRIDA frames the work as three phases~\cite{d71}: anticipate how gaps and technologies may evolve, analyse (the two documents drafted in parallel, each priority checked against the Blueprint) and prepare a roadmap working back from a vision for 2035. It names cross-impact analysis and backcasting among its techniques; its published record is a priority template that cites the Blueprint's gaps, capabilities and actions (D7.1, Annex~B), the evidence tables of Annex~C and a table of dependencies between priorities (D7.1, Table~6.4.1). The evidence has a known scope: the Community of Practice chose the use cases~\cite{d51}, the RA cases predate many SRCNet design decisions~\cite{d61}, consultation respondents self-selected, and theoretical and lattice HEP were under-represented~\cite{expertsmeeting}.

\section{The Technical Blueprint}
\label{sec-tb}

This section follows the Blueprint's four steps: a capability map, consolidated gaps, recommended actions in eight technical areas, and sustainability~\cite{d61}.

\subsection{Capability map}
\label{sec-tb-map}

\begin{figure}[t]
\centering
\includegraphics[width=\textwidth]{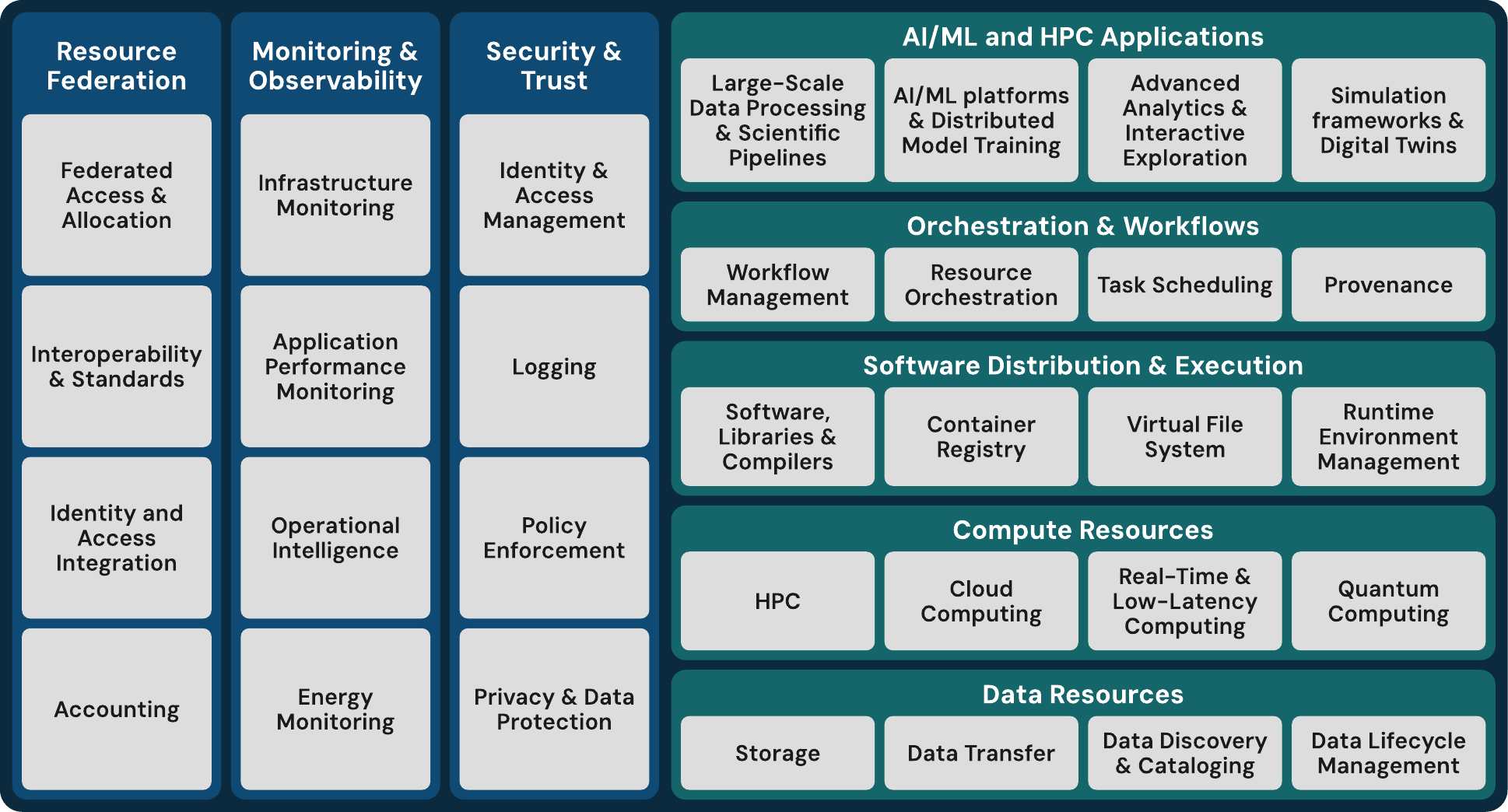}
\caption{Capability map of the European compute and data continuum: five layers (right) and three transversal areas (left). Redrawn from the Technical Blueprint, Figure~1~\cite{d61}, under CC~BY~4.0.}
\label{fig-2}
\end{figure}

The Blueprint treats the continuum as a system of systems linking data centres, HPC, cloud and quantum platforms, networks and instruments, and breaks it down into capabilities: functions or services that the infrastructure provides to scientific workflows~\cite{d61}. Figure~\ref{fig-2} shows the capability map: five layers from resources to applications and three transversal areas.

Most of these capabilities already exist in one or both communities, but they are not necessarily aligned between them or at the same level of maturity. Rucio manages more than an exabyte for ATLAS and CMS across over 120 storage centres and provides data management for SRCNet~\cite{d61,d71}; CernVM-FS distributes software to thousands of WLCG nodes and to LOFAR, and EESSI uses it to deliver optimised stacks to EuroHPC systems~\cite{d61}. Working through the map helped HEP and RA align on these capabilities. For assessments as capabilities mature, SPECTRUM defined a Capability Alignment Maturity Framework with six levels, from unaligned (0) to operational integration (5).

\subsection{Gaps and recommended actions}
\label{sec-tb-gaps}

The Blueprint consolidates sixteen gaps from the use cases, the infrastructure assessment, the access policy analysis and the Community of Practice (D6.1, Table~1). We classify them into four groups. \emph{Access}: fragmented identity, little interactive access, few high-level interfaces, and multi-factor authentication that breaks unattended workflows. \emph{Workflows and software}: stacks that differ between sites and architectures, workflow engines split between communities, and AI/ML platforms detached from data and compute services. \emph{Data}: no unified approach to FAIR storage and provenance, site-specific transfer tools, and near-real-time processing for RA transient searches and some HEP triggers. \emph{Resources}: one- to two-year allocations against decadal programmes, accounting bound to single infrastructures, and no cross-site monitoring. Three gaps cut across all groups: co-design, a fragmented workforce, and security for sensitive data, which HEP and RA primary data do not involve~\cite{d71}.

The recommended actions fall in eight technical areas (D6.1, Sections~5.1 to 5.8), listed in Table~\ref{tab-1} with the SRIDA priorities that take them up. Many have both an organisational and a technical side: strategic access within EuroHPC with a voice in system co-design, joint data challenges, standard HPC interfaces, and a service-account model that HPC security policies can accept. On GPUs and other accelerators the Blueprint is cautious for offline HEP workloads: general-purpose simulation, reconstruction and analysis codes remain largely CPU-bound, pilot ports have so far widened hardware coverage more often than they have delivered large speed-ups, and CPUs will still be needed to drive GPU pipelines~\cite{d61}. Online processing is further ahead: GPU reconstruction runs in production in the Run~3 trigger and synchronous-processing farms of LHCb, ALICE and CMS~\cite{allen,alicels2,patatrack}. RA already runs GPUs at the instrument: LOFAR's correlator and beamformer, COBALT, is GPU-based~\cite{d53}. The planned LOFAR~2.0 fast radio burst search needs 64 GPUs and 1,024 cores with 32~GB of memory per core, and SKA and LOFAR pipelines are expected to move key algorithms to GPUs~\cite{d51}. The portability actions therefore aim to decouple code from hardware.

\begin{table}[t]
\centering
\caption{Recommended action areas of the Technical Blueprint (D6.1, Sections~5.1 to 5.8; abridged) and the SRIDA priorities that take them up, from D7.1 Annex~A and Table~C.4.1 (whose pre-final D6.1 section numbers are matched here by title). P09 and P12 draw on D6.1 Section~3.}
\label{tab-1}
\footnotesize
\begin{tabular}{@{}>{\raggedright\arraybackslash}p{0.2\textwidth}p{0.6\textwidth}>{\raggedright\arraybackslash}p{0.13\textwidth}@{}}
\toprule
Action area & Recommended actions & SRIDA \\
\midrule
5.1 Interface standardisation & Standard HPC access interfaces; middleware interoperable with the EuroHPC Federation Platform (EFP) identity and allocation APIs. & P04 \\
5.2 Co-design of computing, data, networks & Strategic access or long-term allocations in EuroHPC and national HPC, with a voice in system design; programmable networking. & P01, P11, P13 \\
5.3 Software portability & Code surveys for GPU and HPC readiness; portability-framework pilots; CernVM-FS and container distribution; FAIR software curation. & P08, P10 \\
5.4 Data management and networks & Joint SRCNet, WLCG, LOFAR, HPC data challenges; EFP-ready transfer tools (FTS, Rucio, iRODS); aligned site policies. & P05 \\
5.5 Workflow adaptation & Workflow systems integrated with HPC resource managers; common APIs with unattended execution, provenance and orchestration. & P04, P06 \\
5.6 Security and AAI & AARC-compliant AAI mapping VO membership to HPC access; harmonised HPC authorisation policies; auditable service accounts. & P03 \\
5.7 AI/ML integration & Tasks suitable for ML acceleration identified; double-precision needs communicated to HPC planners; ML training. & P07 \\
5.8 Long-term provisioning & Multi-year and strategic HPC projects for allocations aligned with experiment timelines; a study of lifecycle-aligned allocation models. & P01, P02 \\
\bottomrule
\end{tabular}
\end{table}

\subsection{Sustainability}
\label{sec-tb-sust}

The Blueprint addresses sustainability in two senses: preservation and reuse, since archives mostly serve single communities, and environmental impact, assessed over the whole life cycle (ISO~14040 and 14044) across all 16 impact categories, to avoid shifting impacts between phases or categories. For IT equipment, given its short lifetimes and an energy mix such as Europe's, fabrication outweighs use in climate impact, so the first recommendation is to use hardware longer; efficiency gains are not counted as reductions because cheaper resources raise demand~\cite{d61}. SRIDA Priority~12 builds on this and proposes environmental impact, instead of core-hours, as the unit of user credit~\cite{d61,d71}.

\section{The Strategic Research, Innovation and Deployment Agenda}
\label{sec-srida}

\subsection{Drivers, pillars and priorities}
\label{sec-srida-pillars}

The SRIDA starts from seven drivers: digital infrastructure scale, AI adoption, environmental sustainability, security and trust, long-term preservation, workforce capacity and digital sovereignty~\cite{d71}. Its vision for 2035 is a continuum in which researchers run workflows across HPC, high throughput computing (HTC), cloud, quantum and edge resources without concern for infrastructure boundaries, authentication barriers or data locality.

\begin{table}[t]
\centering
\caption{The thirteen SRIDA priorities and their key messages, condensed from D7.1, Section~4~\cite{d71}.}
\label{tab-2}
\footnotesize
\begin{tabular}{@{}>{\raggedright\arraybackslash}p{0.29\textwidth}>{\raggedright\arraybackslash}p{0.67\textwidth}@{}}
\toprule
Priority & Key message \\
\midrule
\multicolumn{2}{@{}l}{\emph{Pillar 1: Policy, Trust and Governance}} \\
P01 Cross-Infrastructure Governance & A coordination forum where WLCG, SRCNet and LOFAR ERIC align and seek formal representation in pan-European digital infrastructures. \\
P02 Federated Resource Allocation & Multi-year allocation aligned with facility timelines and standard accounting, building on WLCG national commitments and EuroHPC Strategic Access. \\
P03 Federated Identity and Trust & Federated access with service accounts for unattended workflows, by making the existing AARC-compliant proxies interoperate~\cite{aarc}. \\
\multicolumn{2}{@{}l}{\emph{Pillar 2: Architecture and Interoperability}} \\
P04 Heterogeneous Computing Integration & Unified, API-based access to HPC, HTC, cloud, edge and quantum resources, so that workflows span providers without custom integration for each one. \\
P05 Data Management and Transport & Cross-domain FAIR data management standards and automated bulk data movement, with remote access for datasets too large to move repeatedly. \\
P06 Workflow Orchestration and Management & Cross-facility orchestration with resilience, provenance capture and unattended execution of long-running workflows. \\
\multicolumn{2}{@{}l}{\emph{Pillar 3: Software and Science Enablers}} \\
P07 AI & AI as first-class infrastructure across HPC, HTC, cloud and edge, with GPU and AI Factory allocations counted as pledged resources. \\
P08 Software Portability & Portable programming models and container distribution, following FAIR4RS~\cite{fair4rs} and building on EESSI. \\
P09 Open Science and Reproducibility & Workflows and results reproducible over five to ten years, through preserved computational context, metadata and provenance. \\
P10 Software Preservation and Sustainability & Scientific codes that stay functional across the 20 to 50 year lifetimes of the instruments, through preservation and modernisation. \\
\multicolumn{2}{@{}l}{\emph{Pillar 4: Human Capital and Responsibility}} \\
P11 Community Collaboration and Co-Design & Channels for community requirements into the coordination forum, building on the Community of Practice and existing networks. \\
P12 Environmental Sustainability & Life-cycle assessment tied to procurement and allocation rules, with environmental impact as the unit of credit given to users. \\
P13 Skills, Careers and Training & Hybrid domain and computing skills through training and career frameworks on the Research Software Engineer model. \\
\bottomrule
\end{tabular}
\end{table}

Table~\ref{tab-2} lists the thirteen priorities with a key message for each. For each pillar the SRIDA also recommends actions for three groups: scientific communities and research infrastructures, e-infrastructures and service providers, and policy makers and funding bodies (D7.1, Sections~4.2.3 to 4.5.3). For access, it asks communities for multi-year resource plans, e-infrastructures to allocate beyond annual cycles, and funders to enable multi-year access.

\subsection{Investment areas and funding instruments}
\label{sec-srida-invest}

The SRIDA maps the priorities to five investment areas: computing infrastructure, data infrastructure, software and tools, governance and coordination, and human capital~\cite{d71}. It also asks funders to support network capacity for exabyte-scale data movement (P05). It estimates that EU framework programmes provide about 10\% of research infrastructure budgets, and it treats national and European funding as complementary: national programmes provide most HTC, cloud and storage capacity, EuroHPC concentrates on exascale and AI-optimised compute, and the task is to align the two streams (P01). The Horizon Europe work programme for 2026--2027 supports cluster and integrated-access activities in the ESFRI domain ``Data, Computing and Digital Research Infrastructures'', which can pay for foundation-phase coordination, software and access but not hardware. For Framework Programme~10 (2028 to 2034, under negotiation), the SRIDA warns that the e-infrastructure federations lose the dedicated line they held under Horizon 2020. It names three costs that budgets underestimate: coordination, software maintenance and the retention of technical staff.

\subsection{Roadmap}
\label{sec-srida-roadmap}

\begin{figure}[t]
\centering
\includegraphics[width=\textwidth]{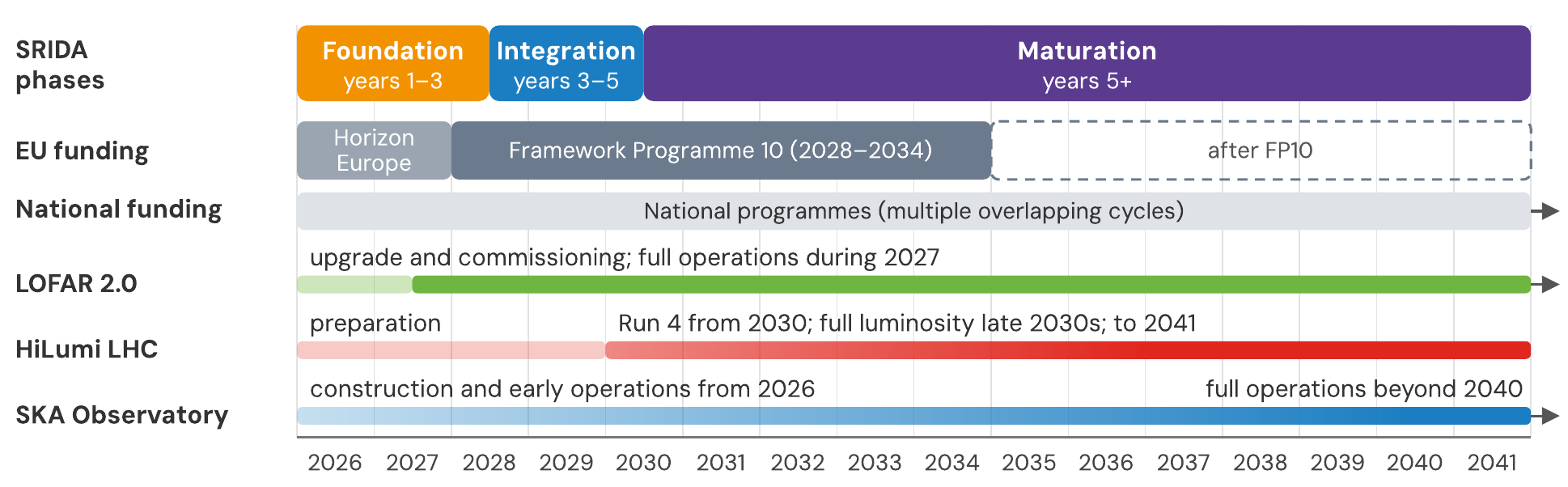}
\caption{The three SRIDA roadmap phases against funding cycles and the timelines of LOFAR~2.0, the HiLumi LHC and the SKA Observatory. Adapted from the SRIDA, Figure~6.2.1~\cite{d71}, under CC~BY~4.0. Phases are counted from 2026, with boundaries in the shared years~3 (2028) and~5 (2030); facility dates as given in D6.1, D7.1 and by CERN~\cite{d61,d71,hilumi}, LOFAR~2.0 as of September 2026. Indicative.}
\label{fig-4}
\end{figure}

The roadmap phases the priorities over three periods (Figure~\ref{fig-4}) under four rules: governance and trust ready before integration at scale, with piloting in parallel; infrastructure and applications scaled up together; investment in people, community and sustainability in every phase; and alignment with EU funding cycles~\cite{d71}. By the end of the foundation phase (2028) a coordination forum, the AARC baseline with cross-infrastructure authentication pilots, and FAIR preservation guidelines should be in place. After the integration phase, workflows span HPC, HTC and cloud with automated data staging, under multi-year allocation agreements (D7.1, Section~6.3). The foundation phase coincides with HiLumi LHC preparation and SKA construction; the integration phase ends as Run~4 begins. The SRIDA asks for operational capabilities by then and production scale before full luminosity in the late 2030s~\cite{d71}.

\section{Conclusion}
\label{sec-conclusion}

The Technical Blueprint and the SRIDA give HEP and RA one account of the compute and data continuum they need: what it must deliver, who should act, when, and with which instruments. The SRIDA's picture is concrete: a researcher reaches reconstructed data across federated sites, runs each step on the resource best suited to it and publishes the results through shared open archives, with one identity, one allocation and one portable software stack~\cite{d71}.

Much of the work is coordination between research infrastructures, digital infrastructures and funders, who could organise the continuum as a commons for research~\cite{commons,osccommons}. Many questions remain open; the two documents set them out as gaps, recommended actions and priorities, and SRIDA Section~7 covers its wider impact~\cite{d61,d71}. We hope they help these actors align their plans for the continuum that HEP, RA and other data-intensive sciences need.

\begin{acknowledgement}
This work is co-funded by SPECTRUM (European Union Grant Agreement Number 101131550). Views and opinions expressed are however those of the authors only and do not necessarily reflect those of the European Union. Neither the European Union nor the granting authority can be held responsible for them. The authors thank the other contributors to and reviewers of both documents, the SPECTRUM Community of Practice, the External Advisory Board, and everyone who gave feedback.
\end{acknowledgement}

\bibliography{refs}

\begin{thebibliography}{24}

\bibitem{hilumi}
{CERN}, {HiLumi LHC} (2026),
  \url{https://home.cern/science/accelerators/hilumi-lhc/}

\bibitem{d61}
{SPECTRUM Project}, {D6.1 Technical Blueprint for Compute and Data Continuum},
  Zenodo (2026), \url{https://doi.org/10.5281/zenodo.20341621}

\bibitem{atlasroadmap}
{ATLAS Collaboration}, {ATLAS Software and Computing HL-LHC Roadmap},
  CERN-LHCC-2022-005 (2022), \url{https://cds.cern.ch/record/2802918}

\bibitem{cmsphase2}
{CMS Offline Software and Computing}, {CMS Phase-2 Computing Model: Update
  Document}, CERN (2022), \url{https://cds.cern.ch/record/2815292}

\bibitem{d51}
{SPECTRUM Project}, {D5.1 Representative Use Cases: Analysis and Alignment},
  Zenodo (2025), \url{https://doi.org/10.5281/zenodo.15310223}

\bibitem{vanhaarlem}
M.P. van Haarlem et~al., {LOFAR: The LOw-Frequency ARray}, Astron. Astrophys.
  \textbf{556}, A2 (2013). \doiwoc{10.1051/0004-6361/201220873}

\bibitem{scaife}
A.M.M. Scaife, {Big telescope, big data: towards exascale with the Square
  Kilometre Array}, Phil. Trans. R. Soc. A \textbf{378}, 20190060 (2020).
  \doiwoc{10.1098/rsta.2019.0060}

\bibitem{d53}
{SPECTRUM Project}, {D5.3 Landscape of Research Infrastructures: Technologies,
  Services, Gaps}, Zenodo (2025), \url{https://doi.org/10.5281/zenodo.15647755}

\bibitem{d71}
{SPECTRUM Project}, {D7.1 Strategic Research, Innovation and Deployment Agenda
  (SRIDA)}, Zenodo (2026), \url{https://doi.org/10.5281/zenodo.21820494}

\bibitem{wlcgstrategy}
D.~Britton, S.~Campana, T.~Boccali, {WLCG Strategy 2024--2027}, WLCG
  Collaboration, Zenodo (2024), \url{https://doi.org/10.5281/zenodo.12623280}

\bibitem{d31}
{SPECTRUM Project}, {D3.1 Community of Practice: Interim Report}, Zenodo
  (2025), \url{https://doi.org/10.5281/zenodo.15100586}

\bibitem{d52}
{SPECTRUM Project}, {D5.2 Interoperable Access Policies: Analysis and
  Recommendations}, Zenodo (2025),
  \url{https://doi.org/10.5281/zenodo.15647608}

\bibitem{hsfcwp}
{HEP Software Foundation}, {A Roadmap for HEP Software and Computing R\&D for
  the 2020s}, Comput. Softw. Big Sci. \textbf{3}, 7 (2019).
  \doiwoc{10.1007/s41781-018-0018-8}

\bibitem{girone2025}
M.~Girone et~al., {Strategy for HPC Integration in WLCG/HEP}, Zenodo (2025),
  \url{https://doi.org/10.5281/zenodo.15032799}

\bibitem{jena2025}
{JENA Computing Initiative}, {JENA White Paper on European Federated
  Computing}, 23 Mar.\ 2025 (2025), \url{https://nupecc.org/jenaa/}

\bibitem{expertsmeeting}
{SPECTRUM Project}, {SPECTRUM Experts Meeting: Summary of Discussions}, Zenodo,
  v2, 26 Mar.\ 2026 (2026), \url{https://doi.org/10.5281/zenodo.19231191}

\bibitem{drafts}
{SPECTRUM Project}, {Draft Technical Blueprint and SRIDA for review}, Zenodo,
  11 Dec.\ 2025 (2025), \url{https://doi.org/10.5281/zenodo.17901519}

\bibitem{allen}
R.~Aaij et~al., {Allen: A High-Level Trigger on GPUs for LHCb}, Comput. Softw.
  Big Sci. \textbf{4}, 7 (2020). \doiwoc{10.1007/s41781-020-00039-7}

\bibitem{alicels2}
{ALICE Collaboration}, {ALICE upgrades during the LHC Long Shutdown 2}, JINST
  \textbf{19}, P05062 (2024). \doiwoc{10.1088/1748-0221/19/05/P05062}

\bibitem{patatrack}
A.~Bocci et~al., {Heterogeneous Reconstruction of Tracks and Primary Vertices
  With the CMS Pixel Tracker}, Front. Big Data \textbf{3}, 601728 (2020).
  \doiwoc{10.3389/fdata.2020.601728}

\bibitem{aarc}
{AARC Community}, {AARC Blueprint Architecture 2025 (AARC-G080)} (2025),
  \url{https://aarc-community.org/guidelines/aarc-g080/}

\bibitem{fair4rs}
M.~Barker et~al., {Introducing the FAIR Principles for research software}, Sci.
  Data \textbf{9}, 622 (2022). \doiwoc{10.1038/s41597-022-01710-x}

\bibitem{commons}
E.~Waegemaekers et~al., {Organizing research as a commons}, SURF report, Zenodo
  (2026), \url{https://doi.org/10.5281/zenodo.21526773}

\bibitem{osccommons}
T.~Ferrari, D.~Scardaci, S.~Andreozzi, {The Open Science Commons for the
  European Research Area}, in \emph{Earth Observation Open Science and
  Innovation} (Springer), pp.~\mbox{43--67} (2018),
  \url{https://doi.org/10.1007/978-3-319-65633-5_3}

\end{thebibliography}

\end{document}